\documentclass[%
prl,
twocolumn,
amsmath,
amssymb,
reprint,
superscriptaddress,
footinbib,
 amsmath,amssymb,
 aps,
longbibliography
]{revtex4-1}
\usepackage[utf8]{inputenc}
\usepackage{lipsum}
\usepackage[english]{babel}
\usepackage{amsmath,amsfonts,amssymb}
\usepackage[T1]{fontenc}
\usepackage{url}
\usepackage{changes}
\usepackage{amsmath}
\usepackage{amsfonts}
\usepackage{amssymb}
\usepackage[version=3]{mhchem}
\usepackage{changes}
\definechangesauthor[color=red]{TJK}
\usepackage{epstopdf}
\usepackage{graphicx}
\usepackage{soul}
\usepackage{changes}
\graphicspath{{./Figures/}}

\begin{document}
\preprint{APS/123-QED}
\title{Heteronuclear soliton molecules in optical microresonators}

\author{Wenle Weng}
\email[]{wenle.weng@epfl.ch}
\affiliation{Institute of Physics, Swiss Federal Institute of Technology Lausanne (EPFL), CH-1015, Switzerland}

\author{Romain Bouchand}
\affiliation{Institute of Physics, Swiss Federal Institute of Technology Lausanne (EPFL), CH-1015, Switzerland}

\author{Erwan Lucas}
\affiliation{Institute of Physics, Swiss Federal Institute of Technology Lausanne (EPFL), CH-1015, Switzerland}

\author{Ewelina Obrzud}
\affiliation{Swiss Center for Electronics and Microtechnology (CSEM), Rue de l'Observatoire 58, 2000 Neuch{\^a}tel, Switzerland}
\affiliation{Geneva Observatory, University of Geneva, Chemin des Maillettes 51, 12901 Versoix, Switzerland}

\author{Tobias Herr}
\affiliation{Swiss Center for Electronics and Microtechnology (CSEM), Rue de l'Observatoire 58, 2000 Neuch{\^a}tel, Switzerland}

\author{Tobias J. Kippenberg}
\email[]{tobias.kippenberg@epfl.ch}
\affiliation{Institute of Physics, Swiss Federal Institute of Technology Lausanne (EPFL), CH-1015, Switzerland}
\begin{abstract}
Optical soliton molecules are bound states of solitons that arise from the balance between attractive and repulsive effects. Having been observed in systems ranging from optical fibers to mode-locked lasers, they provide insights into the fundamental interactions between solitons and the underlying dynamics of the nonlinear systems. Here, we enter the multistability regime of a Kerr microresonator to generate superpositions of distinct soliton states that are pumped at the same optical resonance, and report the discovery of heteronuclear dissipative Kerr soliton molecules. Ultrafast electrooptical sampling reveals the tightly short-range bound nature of such soliton molecules, despite comprising dissipative Kerr solitons of dissimilar amplitudes, durations and carrier frequencies. Besides the significance they hold in resolving soliton dynamics in complex nonlinear systems, such heteronuclear soliton molecules yield coherent frequency combs whose unusual mode structure may find applications in metrology and spectroscopy.
\end{abstract}
\maketitle 

\def \DWRep {\Delta \omega_{\rm rep}}
\def \DFRep {\Delta f_{\rm rep}}

%%%%%%%%%%%%%%%%%%%%%%%%%%%%%%
%%%%%%%%%%%%%%%%%%%%%%%%%%%%%%
\section{Introduction}
%%%%%%%%%%%%%%%%%%%%%%%%%%%%%%

Solitons are one of the most fascinating phenomena in nonlinear dynamics due to their universal spatial or temporal localization of wave forms and the resulting particle-like behavior \cite{akhmediev1997solitons,Stegeman1518}. The ubiquity of solitons has been manifested by the numerous observations in hydraulics \cite{yuen1975nonlinear}, plasmas \cite{bailung2011observation}, lasers \cite{ryczkowski2018real} and Bose-Einstein condensates \cite{khaykovich2002formation,strecker2002formation}, despite the drastic difference in the governing physics. First discovered in optical fibers \cite{kivshar2003optical}, in which case losses play a negligible role, optical temporal solitons have also been observed in ``open systems'', i.e. systems which exhibit dissipation \cite{akhmediev2005dissipative}. In such dissipative systems solitons result from non-equilibrium driving, i.e. from a double balance of loss and gain, as well as dispersion and nonlinearity and correspond to specific solutions of spatiotemporal self-organization. One specific example is dissipative Kerr solitons (DKS), which can form in a continuous wave (CW) driven Kerr nonlinear cavity \cite{leo2010temporal,herr2014temporal}, as mathematically described by the Lugiato-Lefever equation (LLE) \cite{lugiato1987spatial}. With the frequency combs (also referred to as microcombs \cite{kippenberg2018dissipative}) they generate, DKS have been successfully applied in spectroscopy, ranging and telecommunication \cite{dutt2018chip,Obrzud:2019aa,Suh:2019aa,marin2017microresonator,suh2018soliton,trocha2018ultrafast}.

Like their spatial counterparts \cite{stegeman1999optical,crasovan2003soliton,desyatnikov2005optical}, temporal solitons can form bound pairs or groups, akin to molecules. Temporal soliton molecules have been observed in conservative systems such as optical fibers \cite{stratmann2005experimental,hause2008binding,hause2010soliton,hause2013higher}, and have also been theoretically and experimentally investigated in dissipative systems \cite{PhysRevA.44.6954,PhysRevA.64.033814}. Moreover, recent advances in dispersive-Fourier-transformation-based imaging techniques have revealed the formation of soliton molecules in a variety of mode-locked lasers \cite{herink2017real,krupa2017real,liu2018real,he2017supramolecular}. Investigation on soliton molecules provides a direct route to study the interactions between solitary waves, and the formation and dissociation of soliton molecules are closely related to subjects such as soliton collision \cite{roy2005dynamics}, soliton splashing \cite{yi2018imaging}, soliton rains \cite{chouli2010soliton} and the trapping of solitons \cite{jang2015temporal}. Besides the significance they bring to the fundamental understanding of soliton physics, soliton molecules also present the possibility of transferring optical data surpassing the limitation of binary coding \cite{rohrmann2012solitons}.

To date, binding of DKS has only been possible when dispersive waves interlock multiple identical solitons \cite{wang2017universal,cole2017soliton,parra2017interaction}, which leads to the formation of ``homonuclear'' soliton molecules with long-range binding (i.\,e., the distances between solitons are much larger than the widths of solitons). In this work, for the first time we generate \emph{heteronuclear} DKS molecules, which are stable bound states comprised of solitons with \emph{distinct} carrier frequencies, pump detunings, temporal widths and soliton energies. This is achieved by pumping one resonance with a laser that is phase-modulated at a frequency that is only one thousandth of the cavity free spectral range (FSR). This pumping scheme allows us to access a \emph{multistability} regime \cite{hansson2015frequency,anderson2017coexistence} where multiple dynamical microcomb states coexist. Theoretically, besides the usual dispersive-wave-mediated long-range binding, we predict the unusual binding mechanism that results in the direct interaction between dissimilar solitons in close proximity. Experimentally, we apply a dual-comb sampling technique to measure the inter-soliton separation, revealing that distinct solitons can indeed form stable bound structures in systems with instantaneous nonlinear response, despite the fact that the relative phase between constituent solitons is rotating.

%%%%%%%%%%%%%%%%%%%%%%%%%%%%%%
%%%%%%%%%%%%%%%%%%%%%%%%%%%%%%
% =============================================
% =============================================
\begin{figure*} [htp]
\centerline{\includegraphics[width=1.55\columnwidth]{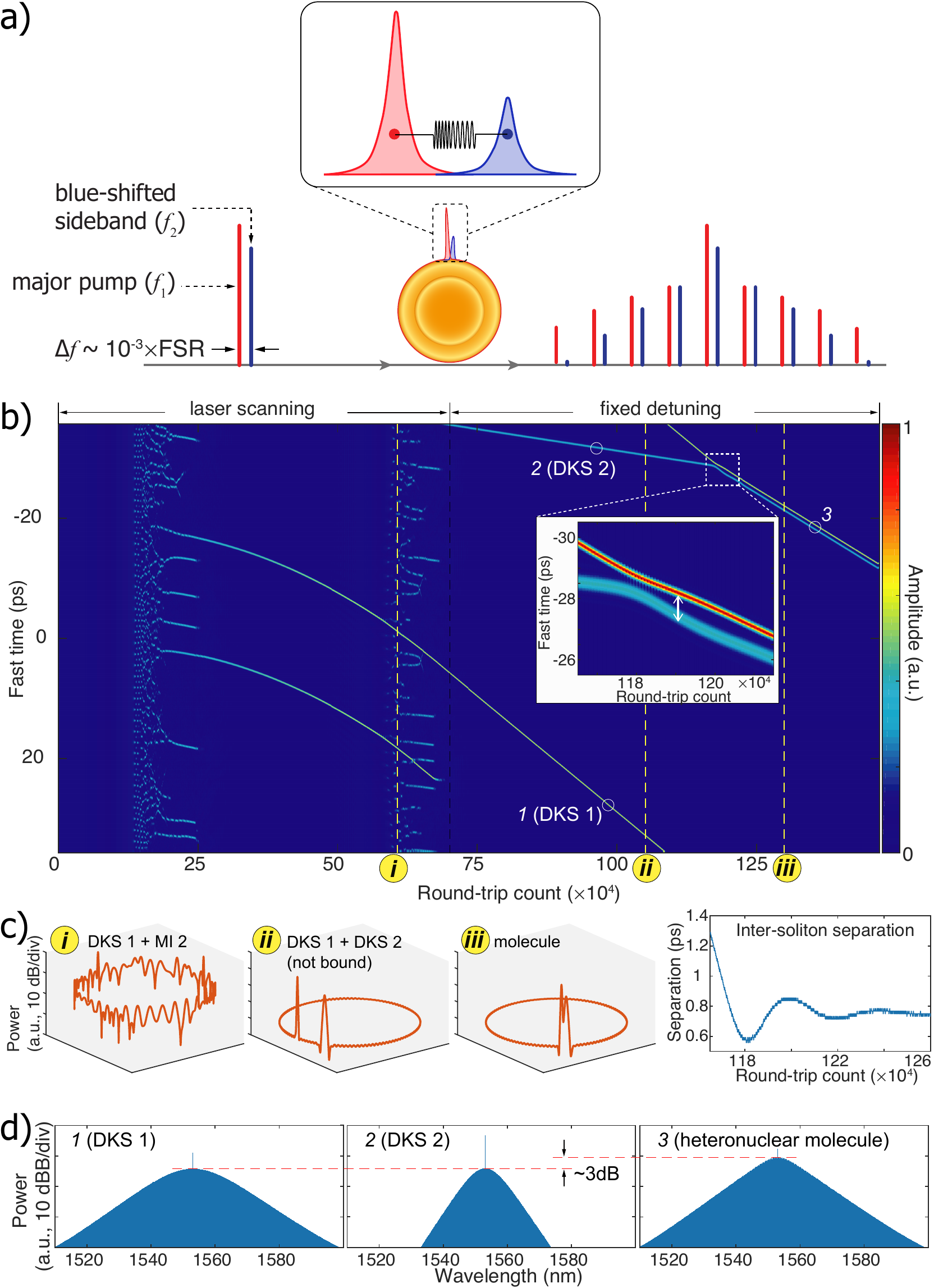}}
\caption{\textbf{Generation of heteronuclear dissipative Kerr soliton molecules.} (a) Principle of the discrete pumping scheme. Closely bound dissimilar solitons are generated by pumping a single resonance with two laser fields of different frequencies simultaneously. (b) Simulated intracavity field evolution showing the formation of soliton molecules. An enlargement of the soliton binding is shown in the inset. Due to the periodicity of the fast time axis the solitons moving out of the map from the bottom will reappear from the top. More details of the simulation can be found in \cite{SM}. (c) Three typical snapshots of the field in (b) at the numbered dashed lines are displayed, corresponding to (1) coexistence of a major soliton and modulational instability; (2) coexistence of unbound dissimilar solitons; and (3) heteronuclear soliton molecule respectively. The evolution of the inter-soliton separation (indicated by the white double-arrow in the inset in (b)) is presented on the right side, showing the stabilization of the separation after oscillations. (d) Frequency comb spectra corresponding to the independent solitons pumped by the major pump (major soliton) and by the sideband (minor soliton) as well as the heteronuclear DKS molecules respectively. Comparison of the comb powers shows that the soliton molecule spectrum is the superposition of the spectra of the two dissimilar solitons.}
\label{fig1}
\end{figure*}
% =============================================
% =============================================

\section{Discretely pumped Kerr microresonators}

%%%%%%%%%%%%%%%%%%%%%%%%%%%%%%
In contrast to a conventional monochromatic pumping scheme, here we drive a \emph{single} resonance with two laser fields, in order to simultaneously generate two distinct soliton states that are each triggered by the bistability of Kerr cavities. Such complex dynamics of multi-valued stationary states, known as multistability, was recently investigated in fiber ring resonators with a single driving laser pumping two resonances \cite{anderson2017coexistence}. Naturally, Kerr microresonators appear to be an ideal platform for studying the multistability since they are more robust against environmental perturbations and their strong nonlinearity allows for wider soliton existence range in the frequency domain. However, the small sizes of microresonators leads to very large FSRs that significantly surpass the bistable range, thus forbidding the multistability regime to be entered. Here we demonstrate that such obstacles can be circumvented by driving one resonance with two laser fields whose frequencies differ by less than the soliton existence range. Since there is only one resonance being pumped, the LLE model is adequate to describe the dynamics. Including a second pump in the driving term, we express the discrete pumping scheme as:

%Recently, it was proposed to extend the bistability of Kerr cavities into the regime of multi-stability to observe multi-valued stationary states, including the coexistence of two soliton states with different detunings driven by a single pump or so-called \emph{super cavity solitons} \cite{hansson2015frequency}. Shortly thereafter, the coexistence of different dynamical states was demonstrated in fiber ring resonators \cite{anderson2017coexistence}, showing the unprecedented rich physics in the regime of multi-stability. Naturally, Kerr microresonators appear to be an ideal candidate for studying the multi-stability since they are more robust against environmental perturbations and their strong nonlinearity allows for wider soliton existence range in the frequency domain. However, the small sizes of microresonators leads to very large FSRs that significantly surpass the bistable range, thus forbidding the multi-stability regime to be entered.

%Here we demonstrate that the multi-stability can be reached by pumping one optical resonance with two laser fields whose frequencies differ by less than the soliton existence range. In this way the pumping fields do not need to be detuned by more than one FSR to let the same mode family to be excited again. Since there is only one resonance being pumped, the LLE model is adequate to describe the dynamics. Including a second pump in the driving term, we express the discrete pumping scheme as:

%
\begin{multline}
\label{eq1}
\frac{\partial{A}}{\partial t} + i \sum_{j=2} \frac{D_j}{j !} (\frac{\partial}{i\partial \phi})^j A - i g {|A|^2 A} =\\ 
\left( { - \frac{\kappa}{2} + i(\omega_0 - \omega_p) } \right){A} + {\sqrt{\kappa _{\rm ex}} \cdot s_{\rm in} (1 +  \frac{\epsilon}{2} e^{i \Omega t})}
\end{multline}
where ${{A}}$ is the envelope of the intracavity field, $\omega_0$ and $\omega_p$ are the angular frequencies of the pumped resonance and the laser respectively, $D_{j}$ is the $j$th-order dispersion, $\phi$ is the co-rotating angular coordinate that is related to the round-trip fast time coordinate $\tau$ by $\phi = \tau \times D_1$ (where $\frac{D_1}{2 \pi}$ is the FSR), ${\kappa}$ is the cavity decay rate, ${\kappa_{\rm ex}}$ is the external coupling rate and ${|s_{\rm in}|^2} = \frac{P_{\rm in}}{\hbar \omega_0}$ is the driving photon flux, where $P_{\rm in}$ denotes the power of the main pump. Here $g =\frac{\hbar\omega_{0}^{2}cn_{2}}{n^{2}V_{0}}$ is the single photon induced Kerr frequency shift, where  $n$ and $n_{2}$ are the refractive and nonlinear optical indices, $V_0$ is the effective mode volume, and $c$ is the speed of light. Moreover, $\epsilon$ is the modulation index for generating the blue-shifted sideband from a phase modulator, e.g.\,an electrooptic modulator (EOM), and $\frac{\Omega}{2 \pi}$ represents the sideband offset frequency, which is set to be positive to reflect the blue-shifted frequency. 

Our theoretical analysis shows that the discrete pumping can simultaneously generate two different soliton states, which can be approximated by the superpositions of solitons excited by only the main pump and by only the sideband, respectively (see Supplemental Material \cite{SM} for details). Here we note that our discrete pumping scheme is fundamentally different from bichromatic pumping methods investigated in previous studies \cite{okawachi2015dual,wang2016dual,ceoldo2016multiple,bao2017dual}. First, while the second pumps in previous works were all offset from the main pump laser by approximately one or multiple FSRs, the offset we apply here is only 12 - 30\,MHz, i.\,e. approximately a thousandth of the FSR (14.09\,GHz). Second, the bichromatic pumping scheme was used for modulating the intracavity CW background to manipulate the spatiotemporal characteristics of the otherwise ordinary monochromatically pumped microcombs. In contrast, the modulated laser in this work generates two dissimilar sets of microcombs, each of which would still exist in the absence of the other's drive.

%%%%%%%%%%%%%%%%%%%%%%%%%%%%%%%%%%%%
\begin{figure*} [t]
\centerline{\includegraphics[width=1.73\columnwidth]{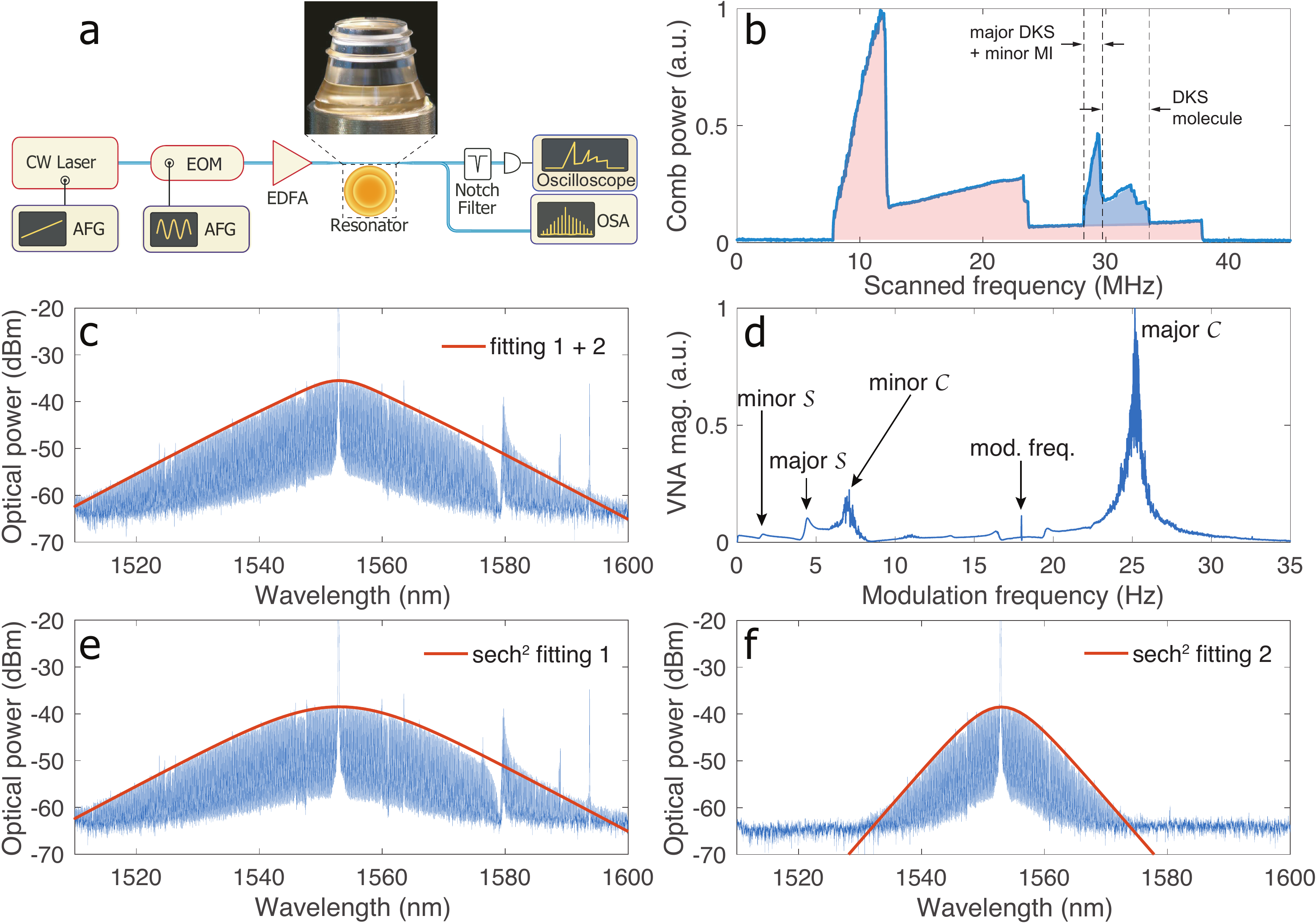}}
\caption{\textbf{Observation of dissipative Kerr soliton molecules.} (a) The schematic experimental setup. (b) The generated comb power as the laser is frequency swept across a resonance. The red-shaded area is the comb power generated by the major pump, while the blue-shaded area indicates the power of the comb driven by the blue-shifted minor pump. (c) The spectrum of the DKS molecule that is composed of major-single-DKS and minor-single-DKS. (d) The measured response of VNA with modulation probing technique. (e) - (f) Spectra of DKS combs driven by only the major pump and by only the minor pump respectively. The red traces are the sech fittings. The sum of the two fittings is plotted in (c).}
\label{fig2}
\end{figure*}
%%%%%%%%%%%%%%%%%%%%%%%%%%%%%%%%%%%(see also the supplementary animation based on simulation \emph{MoleculeFormation.gif})

The simulated formation of heteronuclear soliton molecules is presented in Fig.\,\ref{fig1}\,(b). As the two pumps are swept towards lower frequencies across a resonance, the major pump excites modulational instability (MI) at first, followed by DKS formation. Next, the minor pump scans across the same resonance to generate its microcombs while the major pump is supporting the major DKS. After the minor solitons are formed, due to self-frequency-shifting effects such as high-order dispersion (in this study) and Raman scattering, the major and the minor DKS travel with different group velocities, until the two solitons are close, with a separation where an equilibrium is achieved. Such equilibrium is obtained when the soliton group velocity difference is balanced with the inter-soliton ``repulsion'' caused by cross phase modulation (XPM) (see \cite{SM} for details). Intuitively, in Kerr microresonators where the nonlinear response is instantaneous and local, one may expect that discretely driven solitons can not bind because the Kerr-nonlinearity-mediated effect \cite{hause2008binding} averages out as the relative phase between two distinct solitons is constantly rotating, given that the solitons have different carrier frequencies. Indeed, a fixed phase difference between similar solitons (and their oscillatory tails) is essential to the formation of multi-soliton long-range bound states \cite{wang2017universal,parra2017interaction}. Even in a broader perspective that includes spatial solitons, when the relative phase between driving fields is not fixed (e.\,g. incoherent solitons and vector solitons), bound states can be formed only when the system's nonlinearity has a non-instantaneous or nonlocal nature \cite{mitchell1996self,mitchell1998observation,rotschild2008incoherent}. In our work the XPM effect creates a refractive index barrier to stop solitons of different carrier frequencies (and hence difference group velocities) from colliding. As a result, heteronuclear soliton molecules are formed, with a final group velocity that lies in between the native velocities of the two DKS respectively, according to the conservation of momentum (see \cite{SM} for details).

The corresponding frequency comb spectra are presented in Fig.\,\ref{fig1}\,(d). Because the major and the minor solitons are of different carrier frequencies, the coarsely resolved comb spectra of heteronuclear molecules acquired by an optical spectrum analyzer (OSA) do not show interference patterns that are typical of monochromatically pumped multi-soliton states \cite{brasch2016photonic}. Rather, the averaged comb spectrum of the heteronuclear DKS molecules is the linear superpositions of the spectra of the major and the minor DKS.
%\begin{figure*} [t]
%\centerline{\includegraphics[width=1.8\columnwidth]{fig2.pdf}}
%\caption{\textbf{Observation of experimentally generated dissipative Kerr soliton molecules.} (a) The schematic experimental setup with a picture that shows the MgF$_2$ microresonator. (b) The generated comb power as the laser is frequency swept across an optical mode resonance. The red-shaded area is the comb power generated by the major pump, while the blue-shaded area indicates the power of the comb driven by the blue-shifted minor pump. The comb power generated by the red-shifted sideband produced by the EOM is also presented on the left side of the figure for comparison. (c) The spectrum of the DKS molecule that is composed of major-single-DKS and minor-single-DKS. (d) The measured response of VNA with modulation probing technique. (e) - (f) Optical spectra of DKS combs driven by only the major pump and by only the minor pump respectively. The red traces are the sech fittings for each comb. The sum of the two fittings is plotted in (c).}
%\label{fig2}
%\end{figure*}

% =============================================
\begin{figure*} [htb]
\centerline{\includegraphics[width=1.58\columnwidth]{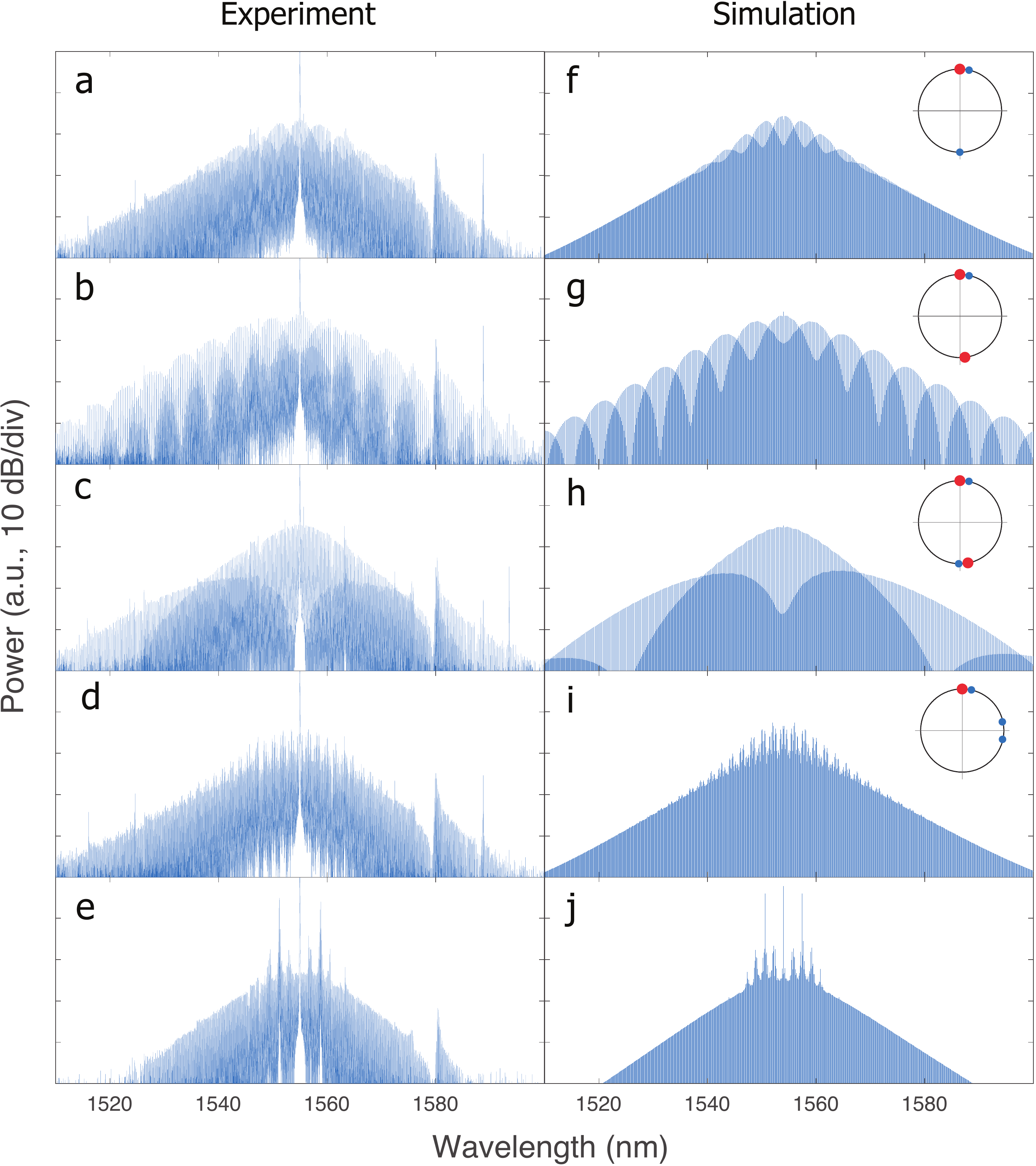}}
\caption{\textbf{Optical spectra of coexistence of distinct solitons.} Experimentally measured (left column) and numerically simulated (right column) spectra of soliton coexistences composed of: (a) and (f) major single-DKS and minor dual-DKS; (b) and (g) major dual-DKS and minor single-DKS; (c) and (h) major dual-DKS and minor dual-DKS; (d) and (i) major single-DKS and minor triple-DKS; (e) and (j) major single-DKS and minor MI. The relative positions of the major solitons (red dots) and the minor solitons (blue dots) are qualitatively depicted on the right side of the simulation panels. For all the simulations the sideband modulation frequency is set to be 18\,MHz and the pump laser detuning is set to be 24\,MHz, except for the superposition of major soliton state and minor MI the pump detuning is 17.7\,MHz.}
\label{fig3}
\end{figure*}
% =============================================

%%%%%%%%%%%%%%%%%%%%%%%%%%%%%%
\section{Experimental generation of heteronuclear soliton molecules}
%%%%%%%%%%%%%%%%%%%%%%%%%%%%%%

We generate heteronuclear DKS molecules in a magnesium fluoride (MgF$_2$) whispering-gallery-mode resonator (WGMR). The experimental setup is depicted in Fig.\,\ref{fig2}\,(a) (see \cite{SM} for more details). Fig.\,\ref{fig2}\,(b) shows the generated microcomb power as the laser frequency sweeps across a resonance. By scanning the laser frequency into the DKS molecule regime, we observe the spectrum of the superposed microcomb of major-single-DKS and minor-single-DKS states (Fig.\,\ref{fig2}\,(c)), while Fig.\,\ref{fig2}\,(e) and (f) show the spectra of the monochromatically pumped single-DKS states driven by the major and the minor pumps respectively. Dynamical probing with vector network analyzer (VNA) \cite{guo2016universal} measures the radio-frequency (RF) spectrum of the transfer function in Fig.\,\ref{fig2}\,(d). We observe two sets of the typical double-resonance features that are induced by the soliton resonance (``$\mathrm{\mathcal{S}}$-resonance'') and the CW resonance (``$\mathrm{\mathcal{C}}$-resonance'') \cite{guo2016universal}, showing that the molecule spectrum is indeed the superposition of two monochromatically driven DKS spectra.

We excite a variety of comb patterns of different compositions, which are presented in Fig.\,\ref{fig3} with the corresponding simulated spectra, showing remarkable agreement. For all the comb patterns we observe only one repetition rate of the out-coupled soliton trains, indicating that the coexistences of solitons are truly bound states. Again, we note that the superposed comb spectra do \emph{not} give information on the separations between the major solitons and the minor solitons in a bound state.

%%%%%%%%%%%%%%%%%%%%%%%%%%%%%%
\section{Structure of soliton molecules}
%%%%%%%%%%%%%%%%%%%%%%%%%%%%%%

Since the observed comb spectrum indicates that the generated solitons are inevitably impacted by high-order dispersion and mode-crossing-induced dispersive waves, the DKS repetition rate ($f_{\rm{rep}}$) depends on the effective pump-resonance detuning \cite{cherenkov2017dissipative, yi2017single}. Consequently, the major and the minor DKS should have different $f_{\rm{rep}}$ because of their different detunings (major detuning $\delta_1 = \frac{\omega_0 - \omega_p}{2 \pi}$ and minor detuning $\delta_2 = \delta_1 - \frac{\Omega}{2 \pi}$) when one assumes that there is no interaction between them. Such assumption, however, is false. Because the major and the minor solitons share the same optical mode family (i.\,e., whispering gallery modes with the same polarization and identical radial and polar numbers), they would either have different $f_{\rm{rep}}$ and show soliton collisions, or have the same $f_{\rm{rep}}$ by forming bound states, i.\,e., heteronuclear DKS molecules. As predicted by our simulation and supported by the single repetition rate of superposed microcombs, the dissimilar solitons travel with same group velocity, likely with a small inter-soliton separation.

\begin{figure*} [t]
\centerline{\includegraphics[width=1.95\columnwidth]{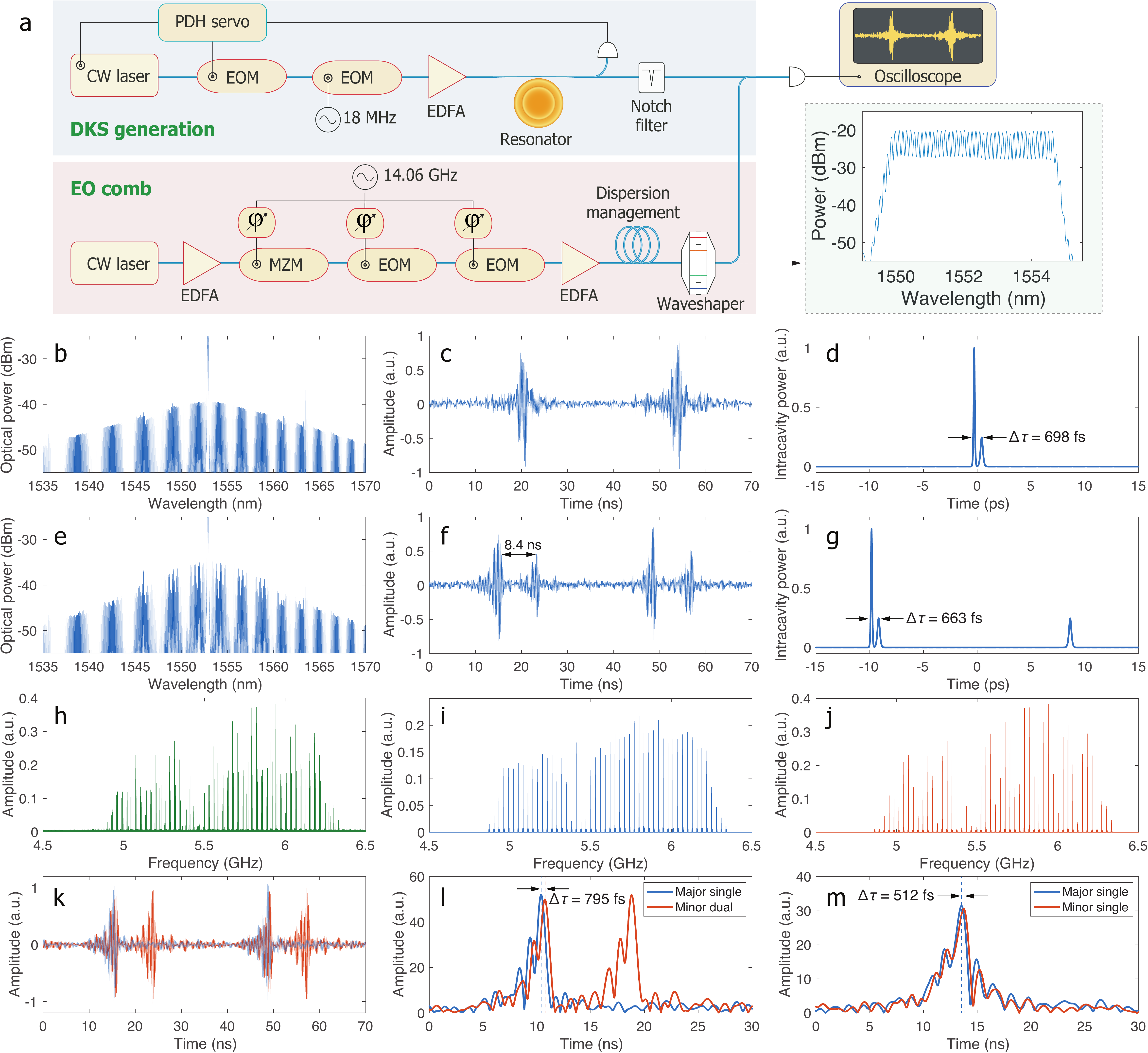}}
\caption{\textbf{Electrooptic examination of the structures of heteronuclear DKS molecules.} (a) The experimental setup and the optical spectrum of the electrooptic comb. The first EOM serves as the phase modulator for generating PDH error signals in order to lock the pump-resonance detuning. (b) - (g) The optical spectrum (b, e), the sampled interferograms (c, f) and the simulated temporal profiles (d, g) of the major-single-with-minor-single and the major-single-with-minor-dual superposed microcombs respectively. (h) - (j) The RF spectrum of the interferograms of the major-single-with-minor-dual DKS molecules and the separated spectra of the major DKS and the minor DKS respectively. (k) The separated interferograms of the microcomb in (e). (l, m) The envelopes (in absolute values) of the separated interferograms of the major-single-with-minor-dual and the major-single-with-minor-single DKS molecules respectively. The inferred real-time separations between the major DKS and the minor DKS are indicated by double-arrows in (d), (g), (l) and (m).}
\label{fig4}
\end{figure*}

%%%%%%%%%%%%%%%
\begin{figure*} [htb]
\centerline{\includegraphics[width=1.7\columnwidth]{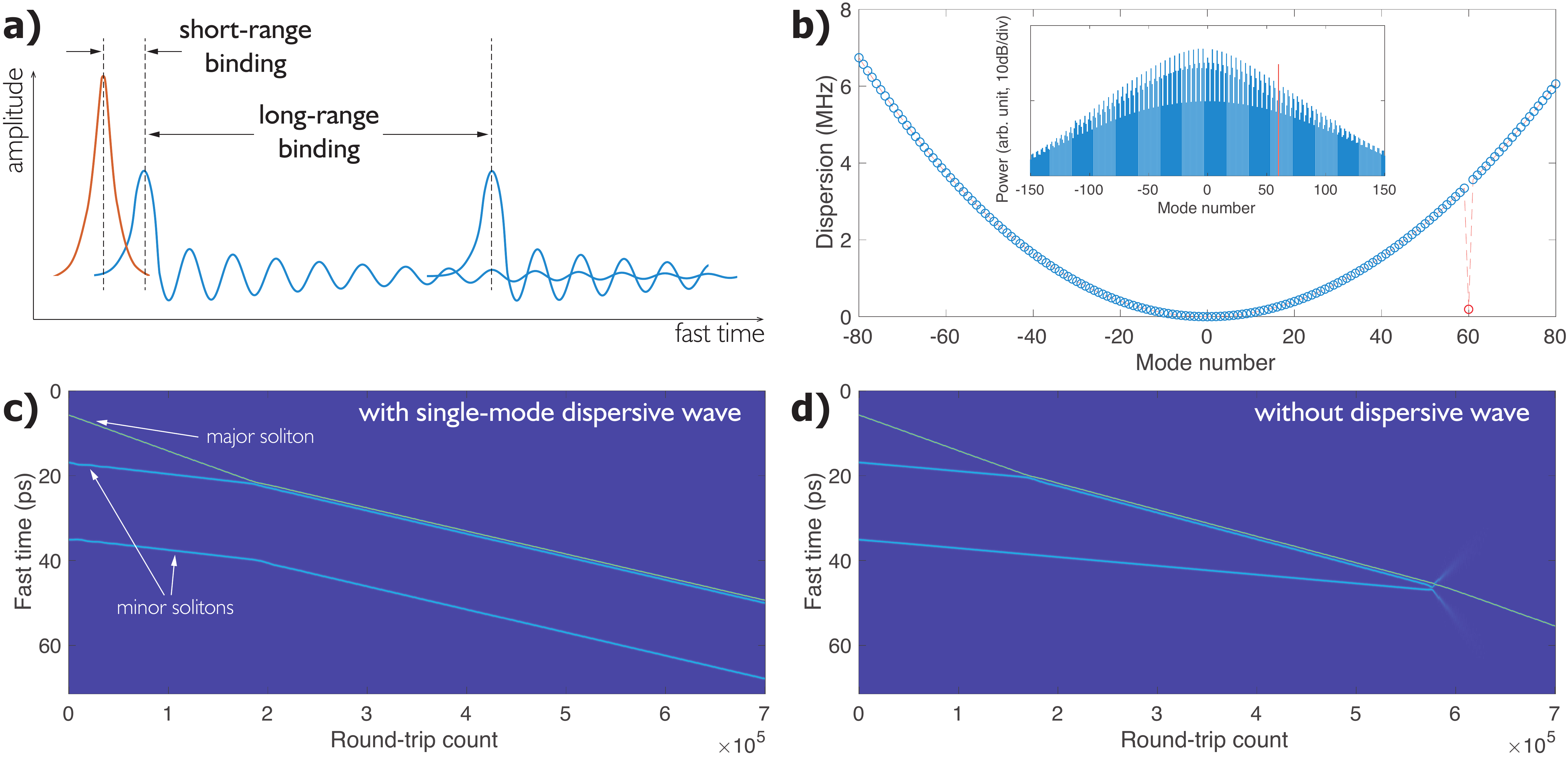}}
\caption{\textbf{Coexistence of long- and short-range binding mechanisms.} (a) Schematic illustration of the coexistence of long-range (dispersive-wave-mediated) and short-range (XPM-induced) binding mechanisms. (b) Dispersion used in the simulation. At mode number of 60 the resonance frequency has a large deviation (shown in red circle) to produce intensive single-mode dispersive wave. The inset displays the comb spectrum, which shows the single-mode dispersive wave in red. (c) Simulated evolution of the bound state of 1 major soliton and 2 minor solitons with the dispersive-wave-mediated binding. (d) Simulated evolution of 1 major soliton and 2 minor solitons without introducing dispersive waves.}
\label{binding}
\end{figure*}
%%%%%%%%%%%%%%%

To verify this prediction, we adopt the imaging technique \cite{yi2018imaging} with an electrooptic comb (EOC) to examine the bound structures. The setup is displayed in Fig.\,\ref{fig4}\,(a). For DKS generation the pump-resonance detuning is stabilized by implementing the Pound-Drever-Hall (PDH) laser locking technique with an EOM as a phase modulator. The major pump frequency is locked to the high-frequency PDH sideband, thus locking the detuning ($\frac{\omega_0 - \omega_{\rm{p}}}{2\pi}$) to be equal to the PDH modulation frequency of 25\,MHz. The carrier frequency of the EOC is different from the soliton pump laser frequency by $\sim5.5$\,GHz. The repetition rate difference ($\Delta f_{\rm{rep}}$) between the EOC and the soliton microcombs is set to be $\sim30$\,MHz. As shown in Fig.\,\ref{fig4}\,(b), (c), (e) and (f), the sampled interferograms show only one repetition period, once again showing that the two constituent microcombs have the same $f_{\rm{rep}}$. However, due to the limited spectral span of the EOC and the chirping of DKS which is introduced by the notch filter, the interferogram duration is typically above 2\,ns, which leads to a temporal resolution that is much broader than the inter-soliton separations between dissimilar DKS, thus potentially forbidding us from uncovering the bound structures. Nevertheless, we use fast Fourier transform (FFT) to transform the sampled interferogram streams of the major-single-with-minor-dual DKS microcomb into the RF spectrum (Fig.\,\ref{fig4}\,(h)). Because the major DKS and the minor DKS have different $f_{\rm{ceo}}$, we are able to decompose the RF spectrum into the major-DKS components (Fig.\,\ref{fig4}\,(i)) and the minor-DKS components (Fig.\,\ref{fig4}\,(j)). Then we apply inverse FFT to transform the separated RF components back into temporal interferograms to infer the temporal delay between the two soliton streams with resolution that is much shorter than the pulse width of the EOC (see \cite{SM} for details).

Fig.\,\ref{fig4}\,(k) shows the separated interferograms of the solitons whose spectrum is presented in Fig.\,\ref{fig4}\,(e). The envelopes of the interferograms are shown in Fig.\,\ref{fig4}\,(l), which also presents the inferred real-time separation between the major and the minor DKS. The same method is also applied to the major-single-with-minor-single soliton streams shown in Fig.\,\ref{fig4}\,(b), and the separated envelopes are plotted in Fig.\,\ref{fig4}\,(m). The derived separations of $\sim500 - 800$\,fs are on the same scale of the temporal pulse width of the individual solitons, providing evidence of extremely short-range binding of distinct solitons.

%%%%%%%%%%%%%%%
\begin{figure*} [t]
\centerline{\includegraphics[width=1.55\columnwidth]{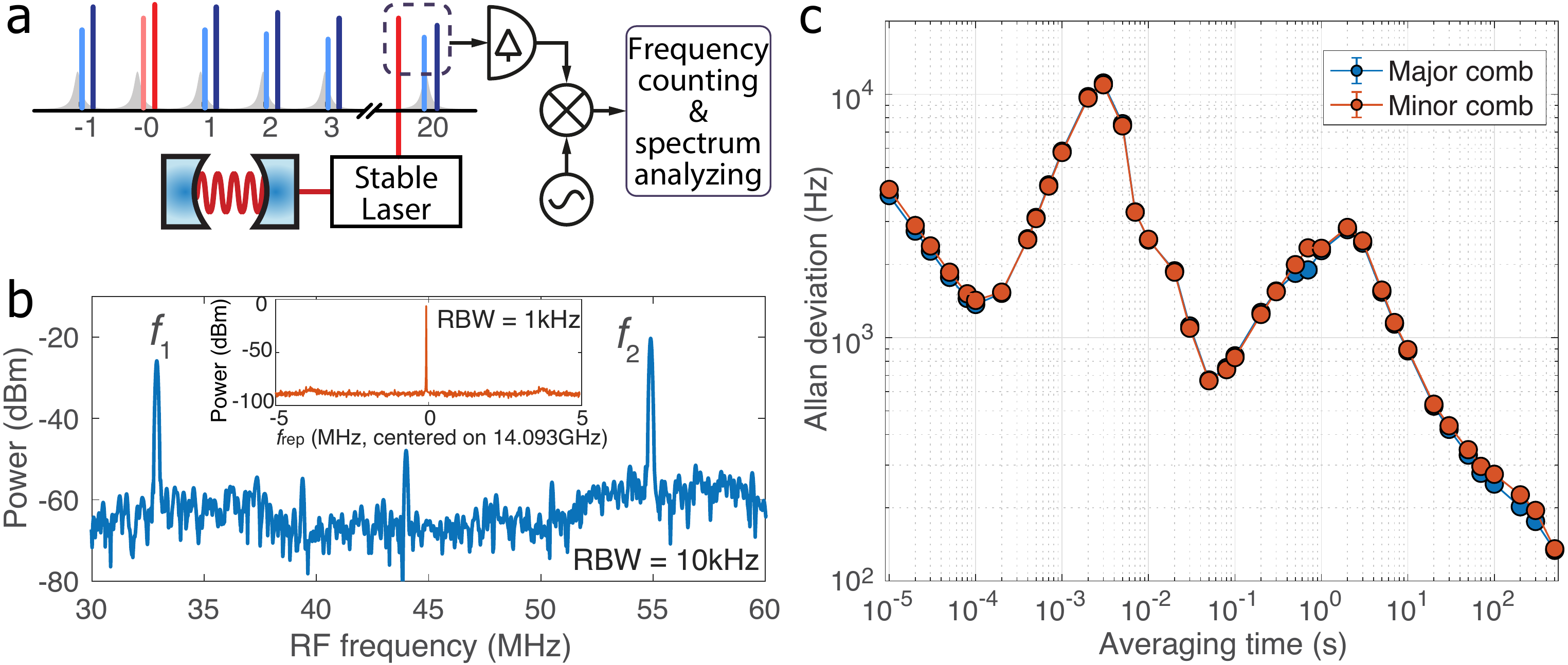}}
\caption{\textbf{Mutual coherence of soliton molecule.} (a) An ultrastable-cavity-locked laser beats with a pair of comb teeth that is 20 FSRs away from the pumped resonance (optical resonances are shown as the gray Lorentzian shapes). The beat frequencies are mixed down to $f_1$ and $f_2$ at low frequency range for frequency counting. (b) The RF spectrum of the mixed-down signals, with resolution bandwidth (RBW) of 10\,kHz. The frequency difference between $f_1$ and $f_2$ is 22\,MHz, which is equal to the sideband offset frequency $\frac{\Omega}{2\pi}$ in the experiment. The weak peak at 44\,MHz is the second order harmonic of the EOM modulation frequency. The inset is the repetition rate spectrum of the soliton molecules, showing only one frequency at 14.093\,GHz. (c) Calculated Allan deviations of stabilized $f_1$ and $f_2$, showing almost identical frequency stabilities. The error bars are included but not visible in the figure due to their extremely small amplitudes.}
\label{coherence}
\end{figure*}
%%%%%%%%%%%%%%%

%%%%%%%%%%%%%%%%%%%%%%%%%%%%%%
\section{Coexistence of short- and long-range binding mechanisms}

To fully understand how complex heteronuclear soliton molecules form, instead of using a single-LLE model, we apply a model of coupled LLEs \cite{SM} to confirm that the XPM effect is responsible for the binding between dissimilar solitons. In Fig.\,\ref{fig4}\,(d) we present the simulated profile of the molecule that is comprised on a single major soliton and a single minor soliton. By including the third-order dispersion in the single-LLE model, we obtain the sub-ps inter-soliton separation, showing good agreement with the experimental result.

For heteronuclear soliton molecules that are of multiple major or minor solitons, the situation is more complicated as the binding between similar solitons relies on interlocking via dispersive waves. Indeed our simulations reveal that the coexistence of the short-range binding due to the XPM effect and the long-range binding caused by the dispersive-wave-mediated effect is essential for the formation of complex molecules beyond the basic form of single major soliton with single minor soliton. Fig.\,\ref{binding}\,(a) depicts the concept of such combined binding mechanism. Fig.\,\ref{binding}\,(b) shows the resonator dispersion used in the qualitative simulation. In order to introduce dispersive wave effect we add large frequency deviation to the resonance with mode number of 60. As can be seen from the inset of Fig.\,\ref{binding}\,(b), such resonance frequency deviation generally leads to enhancement of the comb power in the mode, e.\,g. a single-mode dispersive wave. The simulation is started with a major soliton and two minor solitons seeded in the intracavity field. Fig.\,\ref{binding}\,(c) shows the evolution of the solitons. After the major soliton is bound with one of the minor solitons, the other minor soliton that is well separated from the bound pair also changes its soliton group velocity due to the long-range binding connecting the minor solitons. Consequently the three solitons travel with the same velocity, becoming a complex bound group. To stress the unique role of the dispersive wave, we repeat the simulation without the single mode resonance frequency deviation. As displayed in Fig.\,\ref{binding}\,(d), the simulation for comparison shows that without the long-range binding, the second minor soliton moves with its original velocity after the first minor soliton changes its velocity due to the binding with the major soliton. After some time, the two minor solitons collide into each other, leading to the annihilation of both.

In Fig.\,\ref{fig4}\,(g) we plot the simulated profile of the complex molecule that is corresponding to the experimental observations shown in Fig.\,\ref{fig4}\,(g) and (f). Again, the inter-soliton separation between the major and minor solitons is in excellent agreement with the measured value.

%%%%%%%%%%%%%%%%%%%%%%%%%%%%%%

\section{Frequency coherence measurement}

Despite the frequency offset imposed by the driving lasers, the binding of the solitons mutually locks the repetition rates, thus potentially giving rise to a high frequency coherence between the major and the minor DKS microcombs. To test the coherence, we use a 1553-nm laser whose frequency is stabilized to an ultrastable Fabry-Perot cavity to measure the frequencies of a pair of major and minor comb teeth that is 20 FSRs ($\sim2.3$\,nm) apart from the pumped resonance. The experimental scheme is illustrated in Fig.\,\ref{coherence}\,(a). The RF spectrum of the beat signals (see Fig.\,\ref{coherence}\,(b)) shows two frequencies that differ by the exact value of the EOM modulation frequency which was set to be 22\,MHz in this experiment. We also fully stabilize the microcomb (see SI for details) and then count the two down-mixed beat signals ($f_1$ and $f_2$) at the same time and the recorded frequencies allow us to confirm unambiguously that the frequency of the minor comb is offset from the frequency of the major comb by $\frac{\Omega}{2 \pi}$. The Allan deviations of the two frequencies are displayed in Fig.\,\ref{coherence}\,(c), showing almost identical instabilities. We attribute the imperfect overlap of the Allan deviations to the imperfect synchronization of the counter gating, as well as the fluctuation of the pulse separation in soliton molecules \cite{shi2018observation} and the internal motion of soliton molecules \cite{krupa2017real}. The in-depth analysis of the fluctuations of inter-soliton separations is beyond the scope of this work. Nevertheless, we emphasize here that the frequency coherence the DKS molecule comb exhibited is already sufficient for a wide range of applications in frequency metrology.

%%%%%%%%%%%%%%%%%%%%%%%%%%%%%%
%\section{Discussion}
\section{Conclusion and Discussion}
%%%%%%%%%%%%%%%%%%%%%%%%%%%%%%
We use modulated light to enter a novel multistability regime in a Kerr microresonator to generate heteronulcear soliton molecules. The structures of the soliton molecules, as well as the underlying mechanisms that enable the formation of such new DKS bound states are analyzed experimentally and numerically. The mutual frequency coherence of the generated combs is verified with both spectral analysis and frequency counting.

For practical applications, comb-based sensing and metrology may benefit from heteronuclear DKS molecules that provide an additional coherent comb. In particular, with the feature that the major comb and the minor comb are highly coherent despite the fact that they share no frequency components, the heteronuclear solitons can be used with the interlocking of counter-propagating solitons \cite{Yang:2017aa} to generate ultrahigh coherent dual-comb spectrometer without the overlapping of comb teeth (thus no RF spectrum folding). Furthermore, soliton bound states have been used for optical data buffers \cite{leo2010temporal}, and it has been proposed to use soliton molecules in optical telecommunication to break the restriction of binary coding \cite{rohrmann2012solitons}. Naturally one would expect the DKS molecules to be capable of storing and buffering soliton-molecule-based data.

%It is important to stress that the soliton molecules in dissipative Kerr cavities go beyond the frame of soliton molecules in fibers because the conventional non-inter-soliton ``force'' in conservative systems (i.\,e. the acceleration of solitons due to Raman and high-order dispersion effects) vanishes. In dissipative cavities these effects only shift soliton frequency by a fixed amount. Consequently, the present model based on the balance of forces \cite{hause2008binding,hause2010soliton,hause2013higher} needs to include the cavity periodicity condition to be valid for studying the DKS molecules. We should expect future theoretical studies towards this direction.

\medskip
%%%%%%%%%%%%%%%%%%%%%%%%%%%%%%%%%%%%%%%%%%%%%%%%%%%%%%%%%%%%%%%%%%
%\noindent\textbf{Data availability statement}
%The data used to produce the results of this manuscript will be available on Zenodo upon publication.
%\medskip

%%%%%%%%%%%%%%%%%%%%%%%%%%%%%%%%%%%%%%%%%%%%%%%%%%%%%%%%%%%%%%%%%%
% Bibliography
\noindent\textbf{Authors contributions.} W.W. conceived the concept and the experimental setup, developed theoretical analysis, and performed numerical simulations. W.W. and R.B. performed the experiments and analyzed the data with assistance from E.L.. E.O. and T.H. constructed the electro-optic comb. W.W. wrote the manuscript, with input from other authors. T.J.K. supervised the project.
%\medskip
%%%%%%%%%%%%%%%%%%%%%%%%%%%%%%%%%%%%%%%%%%%%%%%%%%%%%%%%%%%%%%%%%%
%%%%%%%%%%%%%%%%%%%%%%%%%%%%%%%%%%%%%%%%%%%%%%%%%%%%%%%%%%%%%%%%%%

%%%%%%%%%%%%%%%%%%%%%%%%%%%%%%%%%%%%%%%%%%%%%%%%%%%%%%%%%%%%%%%%%%
%%%%%%%%%%%%%%%%%%%%%%%%%%%%%%%%%%%%%%%%%%%%%%%%%%%%%%%%%%%%%%%%%%
\noindent\textbf{Acknowledgments.}
W.W. thanks Hairun Guo for assistance and discussion on numerical simulation. This publication was supported by Contract No. D18AC00032 (DRINQS) from the Defense Advanced Research Projects Agency (DARPA), Defense Sciences Office (DSO), and funding from the Swiss National Science Foundation under grant agreement No.\,163864, No.\,165933 and No.\,176563 (BRIDGE). W.W. acknowledges support by funding from the European Union's Horizon 2020 research and innovation programme under Marie Sklodowska-Curie IF Grant Agreement No. 753749 (SOLISYNTH).
\bibliography{Ref}

% =============================================
% =============================================
% =============================================
\end{document}